\documentclass{aastex}          
\usepackage{spr-astr-addons}    

\newcommand{\ec}{EC\,11481$-$2303}

\newcommand{\logg}{\mbox{$\log g$}}
\newcommand{\Teff}{\mbox{$T_\mathrm{eff}$}}

\begin{document}

\title{\ec\ - A Peculiar Subdwarf OB Star Revisited}

\author{Thomas Rauch}
\and 
\author{Klaus Werner}
\affil{Institute for Astronomy and Astrophysics,
                 Kepler Center for Astro and Particle Physics,
                 Eberhard Karls University,
                 Sand 1,
                 72076 T\"ubingen,
                 Germany
      }
\and
\author{Jeffrey W\@. Kruk}
\affil{Department of Physics and Astronomy, 
                 Johns Hopkins University, 
                 Baltimore, 
                 MD 21218, 
                 U.S.A.
       }

\email{rauch@astro.uni-tuebingen.de}

\begin{abstract}
\ec\ is a peculiar, hot, high-gravity pre-white dwarf. Previous optical
spectroscopy revealed that it is a 
sdOB star with 
\Teff\,=\,41\,790\,K, \logg\,=\,5.84, and He/H\,=\,0.014 by number. 
We present an on-going spectral analysis by means of 
non-LTE model-atmosphere techniques based on 
high-resolution, high-S/N optical (VLT-UVES) and 
ultraviolet (FUSE, IUE) observations.

We are able to reproduce the optical and UV observations simultaneously
with a chemically homogeneous NLTE
model atmosphere with 
a significantly higher effective temperature and lower He abundance
(\Teff\,=\,55\,000\,K, \logg\,=\,5.8, and He\,/\,H\,=\, 0.0025 by number).
While C, N, and O appear less than 0.15 times solar, the iron-group abundance is
strongly enhanced by at least a factor of ten.
\end{abstract}

\keywords{Stars: individual: \ec\ -
          Stars: abundances - 
          Stars: atmospheres - 
          Stars: chemically peculiar -
          subdwarfs
          }

\section{Introduction}
\label{s:introduction}

Subdwarfs of spectral type B and OB (sdB and sdOB stars, respectively) 
represent an extension of 
Horizontal Branch B (HBB) 
stars towards 
higher effective temperatures. 
The sdB stars are found in the range \Teff\,$\approx$\,25\,000\,--\,30\,000\,K. 
The sdOB stars are hotter, up to 40\,000\,K and more,
and they differ spectroscopically from the sdBs by the appearance of lines
of ionized helium, e.g\@. the most prominent He\,{\sc ii}\,$\lambda\,4686$\AA\ line. 
The spectral analysis of sdOB stars is, thus, more reliable because \Teff\ may
be precisely determined from the He\,{\sc i}\,/\,He{\sc ii} ionization equilibrium. 

Metal-abundance determinations in 
Extended Horizontal Branch (EHB) 
stars are of particular interest, not only
because they might provide insight into their evolutionary history, but also
because they may shed light on the question of pulsational instability in these
stars. Several sdB stars define a new instability strip in the 
Hertzsprung-Russell Diagram (HRD)
\citep[see, e.g\@. the review by][]{odon:1999}, 
that was predicted from pulsational models by
\citet{charpinet:1996}. A high iron abundance in subphotospheric layers is
required for pulsation driving.

In an analysis of a large number of EHB stars, \cite{edelmann:2003} detected
that two sdOBs and one sdB are very peculiar. From optical spectra he found
extreme overabundances for many iron-group elements, up to 30\,000 times solar,
and it is thought that the origin is radiative levitation of these elements.

In this paper we examine the case of another peculiar sdOB star, 
\ec\ \citep[][WD\,1148$-$230]{ms:1999}
which was first analyzed by \citet{stys:2000} (Sect.\,\ref{s:discovery}).

\section{Discovery and first spectral analysis}
\label{s:discovery}

\ec\ was discovered in the Edinburgh-Cape Blue Object Survey 
\citep[$V = 11.76, B-V = -0.27, U-B = -1.16$,][]{kilkenny:1997}.
A faint companion was detected at a distance of 6\farcs 6, too far away to have an influence on the evolution of \ec.

A first spectral analysis was performed by \citet{stys:2000}.
Based on
optical ($3350 - 5450$\,\AA, resolution 3.5\,\AA, Aug 15, 1995, 1.9m telescope at SAAO)
and
ultraviolet ($1150 - 1950$\,\AA, resolution 7\,\AA\ (SWP\,48111) and
             0.1\,\AA\ (SWP\,48112), Jul 14, 1993, IUE\footnote{International Ultraviolet Explorer})
observations, they used LTE (Local Thermodynamic Equilibrium)
model atmospheres (provided by Detlev Koester and Pierre Bergeron) that
considered opacities of H and He only.
They tried many model assumptions, 
homogeneously mixed,
chemically stratified,
``spot'' models,
DA + DB binary models,
and weak-wind models
to achieve the best fit to the H Balmer lines H\,$\beta$ to H\,$8$
with a homogeneous, single-star model.
They determined 
\Teff\,=\,41\,790\,K, \logg\,=\,5.84, and He\,/\,H\,=\,0.014 by number. 
With these parameters, no satisfactory fit to the IUE data was possible.
Quite unexpectedly, the observed spectrum is not as steep as it is expected for such a hot subdwarf, 
but instead it is rather flat. 
\citet{stys:2000} speculated on several possible explanations (composite binary spectrum, spot model), 
and concluded that \ec\ resembles the DAB white dwarf GD\,323.
We suggest that extreme line blanketing by strongly overabundant iron-group 
elements may be the true reason for the peculiar UV-flux shape of \ec.

\begin{figure}[ht]
\includegraphics[width=\columnwidth]{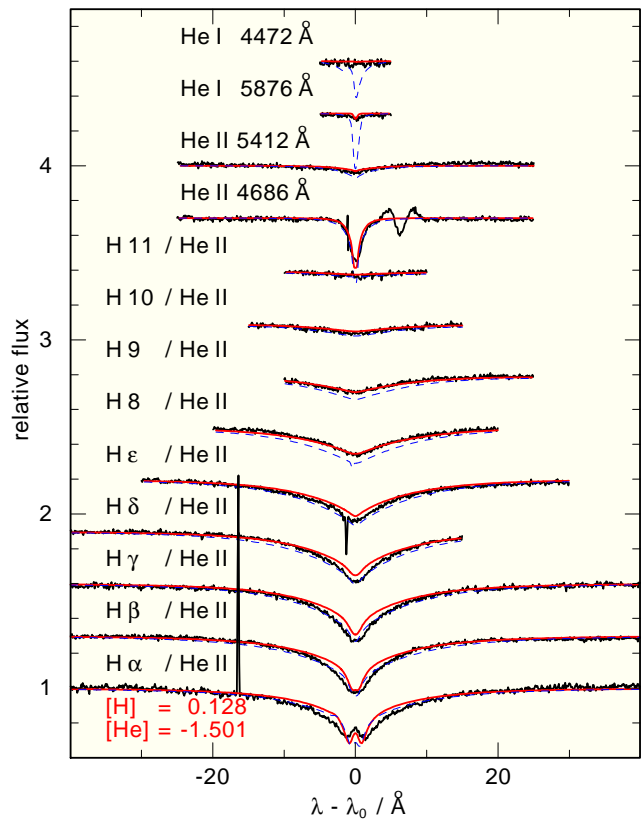}
\caption{Comparison of theoretical H and He line profiles computed from H+He model atmospheres 
         with observations.
         The profiles calculated with parameters of \citet[][Tab.\,\ref{t:preliminary}]{stys:2000} 
         are shown in thin, dashed, blue lines,
         Profiles with our model parameters are overplotted in thick, full, red lines.
         [X] denotes log (mass fraction / solar mass fraction) of species X.
        }
\label{f:stys}
\end{figure}

\section{Preliminary analysis}
\label{s:preliminary}

For our analysis, we used much better observations (re\-solution 0.1\,\AA) in the optical wavelength range 
(3300\,\AA\ $< \lambda <$ 7000\,\AA) which are provided in the framework of the 
ESO\footnote{European Southern Observatory} Supernovae\,Ia Progenitor surveY \citep[SPY,][]{napiwotzki01}
with the VLT\footnote{Very Large Telescope}. 
We used \emph{TMAP}\footnote{http://astro.uni-tuebingen.de/\raisebox{0.3em}{{\tiny $\sim$}}rauch/TMAP.html}, 
the T\"ubingen NLTE Model-Atmosphere Package 
\citep{rd:2003,wea:2003} for the
calculation of plane-parallel model atmospheres in hydrostatic and radiative equilibrium. 
We have to note here, 
that the consideration of H\,{\sc i} line-broadening in the calculation of 
spectral energy distributions (SEDs) 
has changed in \emph{TMAP}.
First, \citet{repolust:2005} found an error in the H\,{\sc i} line-broadening tables (for high members of
the spectral series only) by \citet{lemke:1997} that had been used before by \emph{TMAP}. These 
were substituted by a Holtsmark approximation. Second, \citet{tb:2009} presented new, parameter-free
Stark line-broadening tables for H\,{\sc i} including non-ideal effects. These replace Lemke's data
for the lowest ten members of the Lyman and Balmer series.

In a first step, we use the H and He lines to check for the result of \citet{stys:2000}.
We get significantly different results (Tab.\,\ref{t:preliminary}). This is because,
based on the new observations, we can precisely evaluate the 
He\,{\sc i}\,/\,He\,{\sc ii} ionization equilibrium and then measure the H / He abundance ratio.

\begin{table}[ht!]
\small
\caption{Basic parameters of \ec, derived from H+He model atmospheres.
        }
\label{t:preliminary}
\begin{tabular}{rrr}
\tableline
& \citet{stys:2000} & our work \\
\tableline
\Teff\ / K            & 41\,790 & 55\,000 \\
\logg\ / (cm/sec$^2$) & 5.84    & 5.8     \\
H / He (mass)         & 18      & 100     \\
He / H (number)       & 0.014   & 0.0025  \\
\tableline
\end{tabular}
\end{table}

\begin{figure}[ht]
\includegraphics[width=\columnwidth]{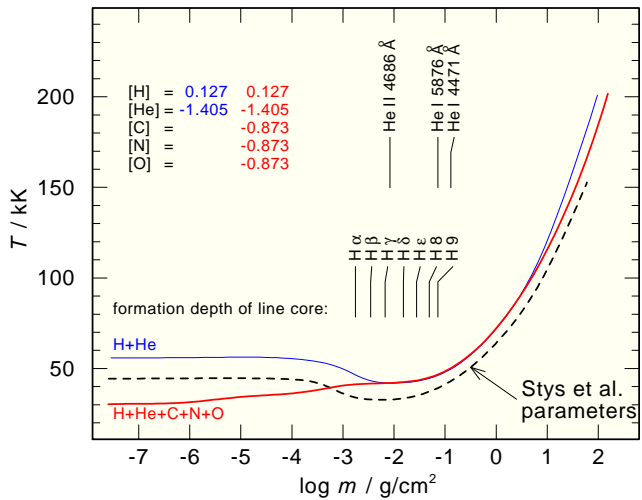}
\caption{Photospheric temperature stratification of H + He models (blue) compared to a H + He + C + N + O models (red).
         The formation depths of H and He line cores are marked.
        }
\label{f:blp}
\end{figure}

In Fig.\,\ref{f:stys}, we show a comparison of spectral lines, calculated with the
parameters from Tab.\,\ref{t:preliminary} with observations. The theoretical line profiles of the lower members 
of the H Balmer series strongly deviate from observations. 
The reason is the so-called
Balmer line problem \citep{werner:1996,bergeron:1993}. The neglection of metal opacities
in the model-atmosphere calculation yields an incorrect temperature stratification in the
line-forming region. Fig.\,\ref{f:blp} shows a comparison of the temperature structure of H+He to
H+He+C+N+O model atmospheres (the optical spectrum provides only upper limits for the C, N, and O
abundances, we adopted these values). The effect is stronger for the lower Balmer-series members.
In Fig.\,\ref{f:blp}, we show also the temperature structure of
a model that was calculated with the parameters of \citet[][Tab.\,\ref{t:preliminary}]{stys:2000}.
Due to the much lower \Teff, its temperature is also much lower in the line-forming regions.
This lets the fits of the Balmer lines calculated from this model appear better than the fits
of our model although the \citet{stys:2000} parameters are unrealistic.

\begin{figure}[ht]
\includegraphics[width=\columnwidth]{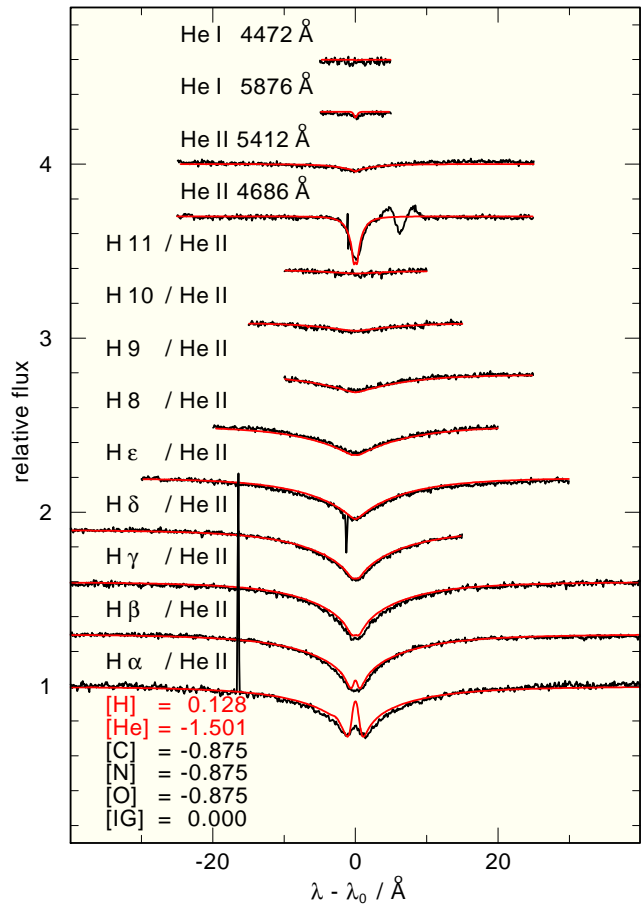}
\caption{Same as Fig.\,\ref{f:stys}. In our models we consider in addition opacities of C, N, O, 
         and of the iron-group (IG) elements.
        }
\label{f:metal}
\end{figure}

\begin{figure}[ht]
\includegraphics[width=\columnwidth]{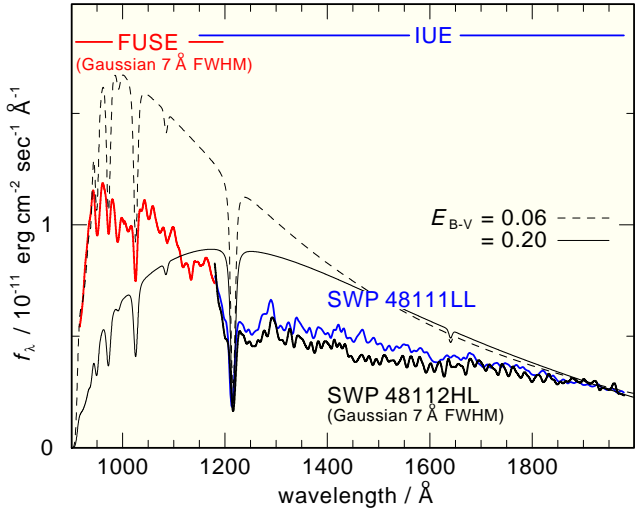}
\caption{\emph{TMAP} SEDs with two different values of $E_\mathrm{B-V}$ 
         \citep[reddening law of][$R_\mathrm{V}=3.2$]{fitzpatrick:1999}
         compared to the FUSE and IUE observations of \ec.
         The reddening ($E_\mathrm{B-V}=0.06$)
         and the interstellar H\,{\sc i} column density ($3\cdot 10^\mathrm{20}$\,cm$^{-2}$) 
         are adopted from \citet{stys:2000}. 
         All SEDs are normalized to match the observed flux at 1950\,\AA. The FUSE and IUE high-resolution
         observations are convolved with a Gaussian in order to match the resolution of the IUE low-resolution
         observation.
        }
\label{f:uv}
\end{figure}

In order to demonstrate the impact of metal opacities, we include C, N, O, and the iron group (IG, here
Ca -- Ni) in our \emph{TMAP} models. For C, N, and O, we used the upper abundance limits (see above),
for IG we assume a solar mass fraction \citep{ags:2005}. In Fig.\,\ref{f:metal}, we compare theoretical and
observed line profiles of H\,{\sc i}, He\,{\sc i}, and He\,{\sc ii} in the optical wavelength range.
The agreement of the line wings of the lower members of the H Balmer series is much better compared to the
H+He models (Fig.\,\ref{f:blp}).

We can summarize, that the optical spectrum of \ec\ can be reproduced by a chemically not stratified NLTE
model atmosphere with \Teff\,=\,$55\,000 \pm 5\,000$\,K, \logg\,=\,$5.8 \pm 0.3$, and H\,/\,He\,=\,$100 \pm 0.3$\,dex (by mass)
when metal additional opacities are considered. Since the optical spectrum does not provide further
information about metal opacities, we will continue with an analysis of the UV spectrum of \ec\ in the
following section.

\section{UV: observations and spectral analysis}
\label{s:uv}

In addition to the IUE observations that were already used by \citet{stys:2000},
we use FUSE\footnote{Far Ultraviolet Spectroscopic Explorer} observations (May 21, 2001, obs id B0540901).
They were performed in four groups (exposure times $432 - 567$\,sec) that
were separated by about one FUSE orbital period ($\approx$\,6000\,sec).
The total exposure time is 8300\,sec. The S/N ratio is $30 - 50$ per pixel at a
resolution of $\approx 0.07$\,\AA. We cannot detect any systematic flux variation in these four
groups. Thus, there is no direct hint for a companion star.

The UV observations are shown in Fig.\,\ref{f:uv}. Their flux levels match well.
The comparison to \emph{TMAP} SEDs shows clearly, that reddening cannot be the only reason 
that the UV flux appears much flatter than predicted by a H+He model.

\begin{figure}[ht]
\includegraphics[width=\columnwidth]{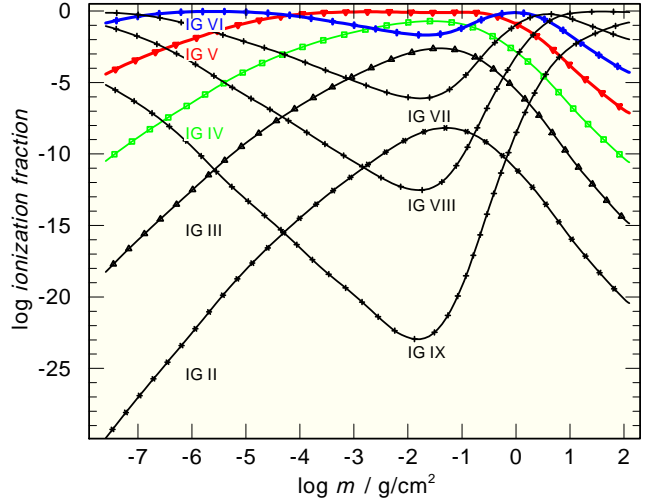}
\caption{Ionization fraction of the generic iron-group element in our model atmosphere with
         \Teff\,=\,$55\,000\,K, \logg\,=\,$5.8, 
          [H]\,=\,0.127,
          [He]\,=\,$-$1.405,
          [C]\,=\,$-$0.875,
          [N]\,=\,$-$0.875,
          [O]\,=\,$-$0.875, and
          [IG]\,=\,0.000.
        }
\label{f:ionfrac}
\end{figure}

At the higher \Teff\ derived by us, radiative levitation could be very efficient.
This supports our idea that line blanketing of overabundant iron-group 
elements is responsible for the flat UV flux. We calculated model atmospheres
with different iron-group abundances ([IG]\,=\,0, 1, 2). For these calculations, the complete
iron group is represented by a generic model atom which was created by our
\emph{IrOnIc} code \citep{rd:2003}. Fig.\,\ref{f:ionfrac} shows the ionization fractions
of the generic iron-group element (IG). IG\,{\sc iv} to IG\,{\sc vi} are the dominant 
ionization stages in the line-forming region in the relevant parameter range.
In our calculations, we consider IG\,{\sc ii} - {\sc ix}.

Our main data source for iron-group elements are Kurucz' data files \citep{kurucz:1997}.
For line transitions, these are available as so-called ``POS'' lists, that include lines
with laboratory measured ``good'' wavelengths, and ``LIN'' lists, that include 
theoretically calculated wavelengths in addition. \emph{TMAP} uses the LIN lists
for model-atmosphere calculations in order to have a reliable total opacity.
For a comparison with observations, in general POS line lists are applied.
For our comparison of the observed UV flux of \ec, however, we have to use
LIN lists because due to the much lower number of lines in the POS lists
(e.g\@. for Fe\,{\sc vi} is the number ratio LIN\,:\,POS\,=475\,750\,:\,1100)
results in an unrealistically high flux level.

\begin{figure}[ht!]
\includegraphics[width=\columnwidth]{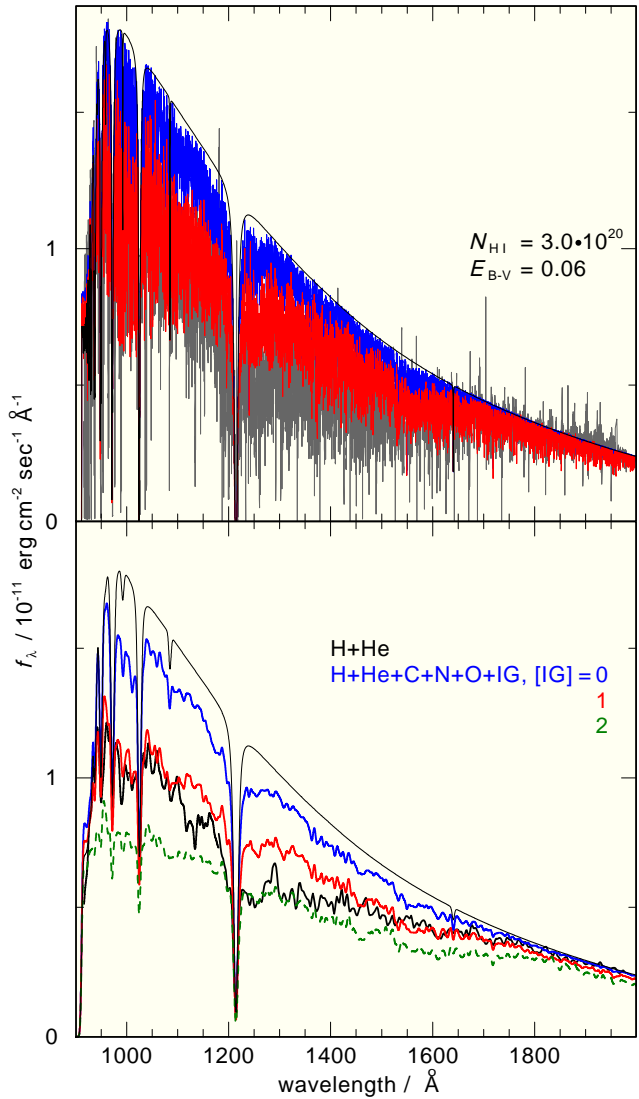}
\caption{SEDs of our final model atmospheres (parameters see Fig.\,\ref{f:ionfrac}) with
         different iron-group abundances compared with observations. 
         At bottom, the SEDs and the FUSE observation are
         convolved with a Gaussian of 5\,\AA\ (FWHM) for clarity.
         In the top panel, one can get an impression of the extremely high number
         of spectral lines that are considered.
        }
\label{f:ig}
\end{figure}

In Fig.\,\ref{f:ig}, we compare theoretical SEDs calculated with different iron-group 
abundances with observations. At a ten times solar abundance, the agreement with
observations is good except for $1100\,\mathrm{\AA}\ < \lambda < 1420\,\mathrm{\AA}$.
In this wavelength interval, the theoretically predicted flux is still about 20\,\% higher
than observed. Two possible reasons may be that the individual iron-group elements are differently
enhanced by radiative levitation and that there are still not enough lines considered.

\begin{figure}[ht]
\includegraphics[width=\columnwidth]{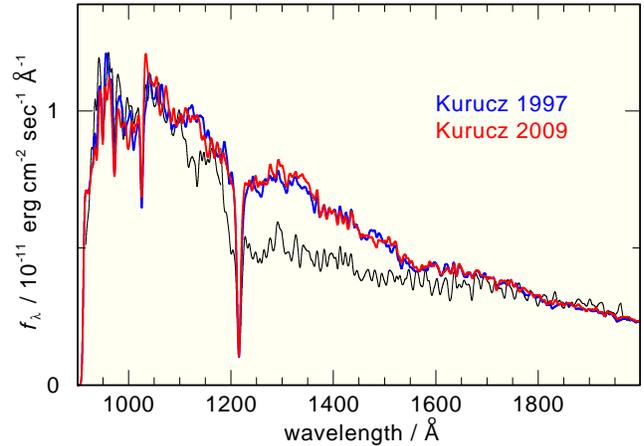}\vspace{-2mm}
\caption{Comparison of two SEDs of our [IG]\,=\,1 model, calculated with
         the old and the extended LIN line lists.
        }
\label{f:kurucz}
\end{figure}

One of the main challenges for the spectral analysis of \ec\ is how to include all the
lines in the \emph{TMAP} calculations, i.e\@. where to get all the data from.
Kurucz' data files were recently extended \citep{kurucz:2009} 
but mainly for the lowest ionization stages. 
The only exception is Fe where new data files are available up to Fe\,{\sc vi}.
For  Fe\,{\sc v}, e.g., the number of LIN lines increased from 1\,000\,385 to 7\,785\,320.
We calculated a new model with the extended LIN lists.
Although many more lines are considered on an even more refined frequency grid,
the differences are within a few percent only (Fig.\,\ref{f:kurucz}).

\begin{figure}[ht]
\includegraphics[width=\columnwidth]{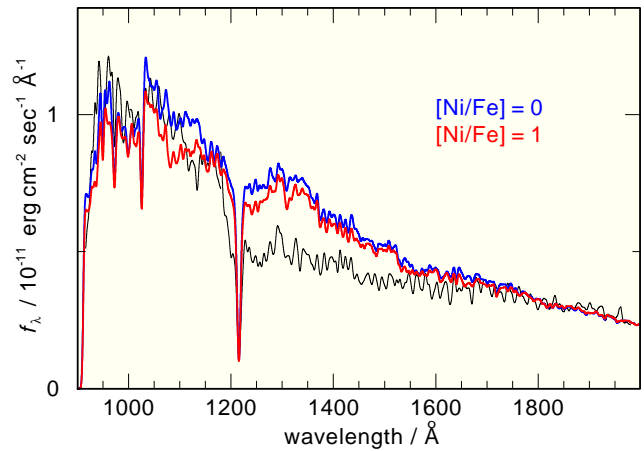}\vspace{-2mm}
\caption{Comparison of two SEDs of our [IG]\,=\,1 model 
         (calculated with the extended LIN line lists), 
         with two different Ni/Fe abundance ratios, 
         compared with observations.
        }
\label{f:nife}
\end{figure}

Changes in the individual abundances of the iron-group elements result in
stronger changes in the UV flux. We performed a test calculation and increased the
Ni abundance in the abundance solar pattern of our generic IG model atom 
by a factor of ten (Fig.\,\ref{f:nife}). The flux is reduced in sections of the
UV spectrum where Ni lines dominate. 

\begin{figure*}[ht]
\includegraphics[width=\textwidth]{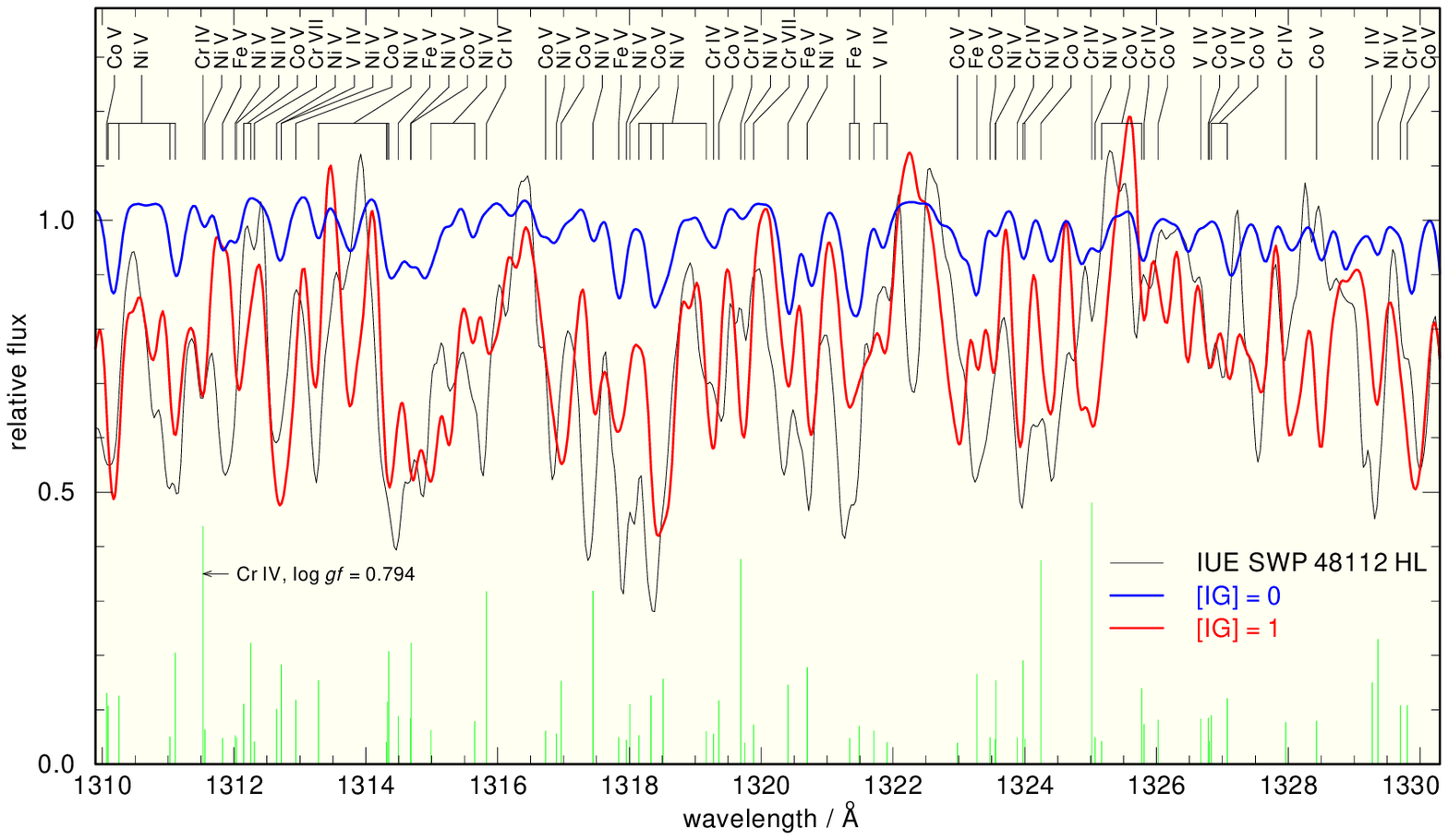}
\caption{Comparison of two synthetic spectra with different iron-group abundances
         with the high-resolution IUE observation. 
         The strongest line transitions ($\log gf \ge -1$) from Kurucz's POS lists are marked at the top. 
         Their $\log gf$ values are indicated at the bottom by vertical bars.
        }
\label{f:iue}
\end{figure*}

We conclude that fine tuning of all
iron-group elements is necessary in order to achieve a better agreement with observations. 
This is corroborated by a comparison of
two of our model SEDs, calculated with [IG]\,=\,0 and [IG]\,=\,1 and the LIN line lists
with the high-resolution IUE observation (Fig.\,\ref{f:iue}). The higher IG abundances
yield a much better agreement with observations. 
A close look shows that individual lines can be identified, e.g\@.
Co\,{\sc V} $\lambda\lambda\,1310.07, 1329.80$\,\AA, which agree well with observations
and will allow a precise abundance determination.
On the other hand, some strong features appear in the synthetic SEDs but 
they are not marked, e.g\@. around 1328\,\AA. These are found only in the LIN line lists 
and their true wavelength positions may be far off.

\begin{figure*}[htb]
\includegraphics[width=\textwidth]{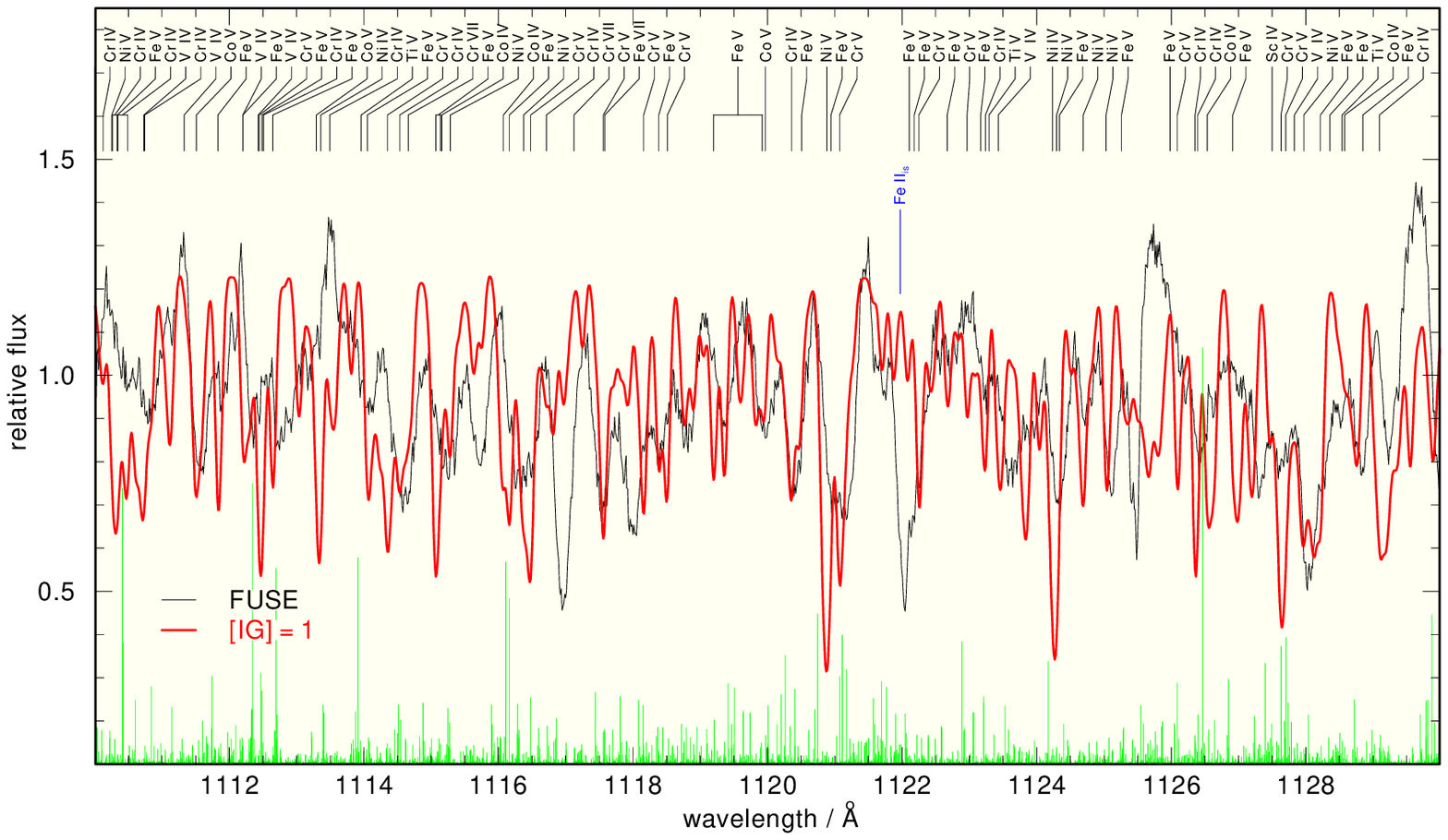}
\caption{Same as Fig.\,\ref {f:iue}, but for a [IG]\,=\,1 model and the FUSE observation.
         At the bottom, the $\log gf$ values of the strongest line transitions ($\log gf \ge -1$) 
         from Kurucz's LIN lists are indicated.
         ``is'' denotes interstellar lines.
        }
\label{f:fuse}
\end{figure*}

Fig.\,\ref{f:fuse} compares a theoretical [IG]\,=\,1 SED with the FUSE observation.
Note the excessively large difference in the numbers of POS and LIN lines that are
marked at the top and bottom, respectively.
As well as in Fig.\,\ref{f:iue}, we see an indication that fine tuning of iron-group
abundances is necessary. Some prominent absorption features, e.g\@. at $\lambda\,1117$\,\AA\
and $\lambda\,1122$\,\AA, appear not reproduced by our model. These might either
stem from the iron group but then, they are not in the LIN lists or their wavelengths are uncertain
and they appear at a wrong wavelength, or they are of interstellar origin (in case of
$\lambda\,1122$\,\AA, Fe\,{\sc ii} is a good candidate for an identification).
\citet{rauch:2007} did show that a simultaneous fit of both, stellar spectrum as well as
interstellar line spectrum, allows to identify isolated, unblended stellar lines and to improve
both, the ISM model as well as the photospheric model. In the further course of our \ec\
analysis, we will include the interstellar lines in our modeling.

\clearpage

\section{Results and conclusions}
\label{s:results}

The optical and UV observation of \ec\ 
are reproduced by our \emph{TMAP} models with
\Teff\,=\,$55\,000 \pm 5000$\,K, 
\logg\,=\,$5.8 \pm 0.3$, 
[H]\,=\,0.127,
[He]\,=\,$-$1.405,
[C]\,$<$\,$-$0.875,
[N]\,$<$\,$-$0.875,
[O]\,$<$\,$-$0.875, and
[IG]\,$>$\,1.
C, N, and O abundances are upper limits only, determined from the optical spectrum.
Our calculations show that Ni is probably as abundant as Fe.
The error in the H and He abundances is about 0.3\,dex.

\begin{figure}[ht!]
\includegraphics[width=\columnwidth]{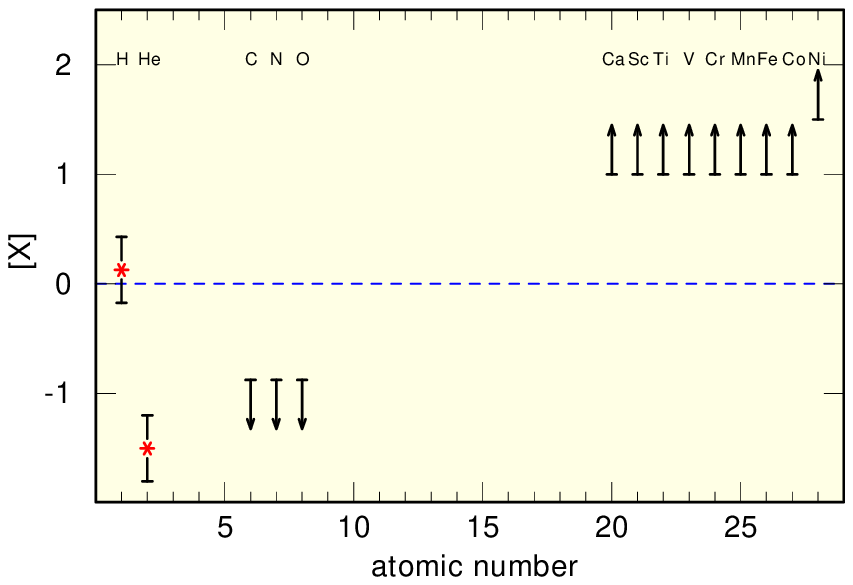}
\caption{Photospheric abundances of \ec.\vspace{7.2mm}
        }
\label{f:abund}
\end{figure}

Our suggestion is that the photospheric abundances of \ec\ (Fig.\,\ref{f:abund}) display
the interplay of gravitational settling and radiative levitation.
The latter is responsible for the strong iron-group over-abundance.

\ec\ and other sdOB and sdB stars with such a high iron-group abundance
\citep[UVO\,0512$-$08, PG\,0909+276, and UVO\,1758+36,][]{edelmann:2003}
should be subjects to detailed diffusion calculations.

There are still deviations between the synthetic spectrum and observations (Fig.\,\ref{f:ig},
$1100\,\mathrm{\AA}\ < \lambda < 1420\,\mathrm{\AA}$). Most likely, these 
are due to our assumption of a solar abundance pattern within our representation of 
Ca -- Ni by one generic model atom. A detailed spectral analysis with individual model
atoms for all iron-group elements is still on-going. Within the framework of this analysis, 
a precise determination of the reddening (including infrared measurements like e.g\@.
2MASS) and the interstellar line absorption will be performed.

The extension of Kurucz's data files is highly desirable because we are strongly
hampered in the spectral analysis of hot stars by the lack of reliable atomic data of 
the expected high ionization stages. This is a challenge for the near future.
\vspace{10mm}

\acknowledgments
T.R\@. is supported by the German Aerospace Center (DLR) under grant 05\,OR 0806.
J.W.K\@. is supported by the FUSE project, funded by NASA contract NAS532985.
This research has made use of the SIMBAD Astronomical Database, operated at CDS, Strasbourg, France.

\bibliographystyle{spr-mp-nameyear-cnd}
\bibliography{rauch.bbl}

\end{document}